\documentclass[utf8]{frontiersSCNS} 

\usepackage{url,hyperref,lineno,microtype,subcaption}
\usepackage[onehalfspacing]{setspace}
\usepackage{gensymb}

\def\keyFont{\fontsize{8}{11}\helveticabold }
\def\firstAuthorLast{Ramlugun {et~al.}} 
\def\Authors{Girish S. Ramlugun\,$^{1}$, Belvin Thomas\,$^{1}$, Vadim
  N. Biktashev\,$^{2,3,4}$, Diane P. Fraser\,$^{2,3}$, Ian J. LeGrice\,$^{1}$, Bruce H. Smaill\,$^{1}$, Jichao Zhao\,$^{1}$, and Irina V.~Biktasheva\,$^{2,4,5,*}$}
\def\Address{$^{1}$ Auckland Bioengineering Institute, The University
  of Auckland, New Zealand \\
$^{2}$CEMPS, University of Exeter, Exeter, UK \\
$^{3}$EPSRC Centre for Predictive Modelling in Healthcare, University of Exeter, Exeter, UK \\
$^{4}$Kavli Institute for Theoretical Physics, UC Santa Barbara, USA \\
$^{5}$Department of Computer Science, University of Liverpool, Liverpool, UK }
\def\corrAuthor{Dr Irina V.~Biktasheva 
}

\def\corrEmail{ivb@liverpool.ac.uk}

\usepackage{graphicx}
\usepackage{dcolumn}
\usepackage{bm}

\usepackage{verbatim}
\newcommand{\code}[1]{\texttt{#1}}

\newcommand{\bbx}{BeatBox}
\newcommand{\bbg}{\code{.bbg}} 

\newcommand\pd{\partial}		

\newcommand{\df}[2]{\frac{\pd{#1}}{\pd{#2}}} 

\def\eqreftwo(#1,#2){(\ref{eq:#1},\ref{eq:#2})}

\newcommand{\kron}{\delta}              

\newcommand{\mx}[1]{\mathbf{#1}}        

\newcommand{\Real}{\mathbb{R}}          

\newcommand{\+}[2]{\def#1{{#2}}}
\newcommand{\1}[2]{\def#1##1{{#2}}}

\+{\num}{^{\sharp}}           
\+{\Boundary}{\Gamma}         
\+{\C}{C}                     
\+{\CM}{C_m}                  
\+{\chf}{\psi}                
\+{\coeff}{\nu}               
\+{\D}{D}                     
\+{\Dten}{\hat{D}}            
\+{\Deten}{\hat{D}^e}         
\+{\Diten}{\hat{D}^i}         
\+{\Ppar}{P_{\parallel}}      
\+{\Port}{P_{\perp}}          
\+{\Depar}{D^e_{\parallel}}   
\+{\Deort}{D^e_{\perp}}       
\+{\Dipar}{D^i_{\parallel}}   
\+{\Diort}{D^i_{\perp}}       
\+{\Deeff}{D^e_{*}}           
\+{\Dieff}{D^i_{*}}           
\+{\Deff}{D_{*}}              
\+{\Domain}{\mathcal{D}}      
\+{\di}{\Delta_i}             
\+{\dj}{\Delta_j}             
\+{\dk}{\Delta_k}             
\+{\dt}{k}                    
\+{\dx}{h}                    
\1{\E}{\times10^{#1}}         
\+{\err}{\varepsilon}         
\+{\f}{\mx{f}}                
\+{\fib}{f}                   
\+{\fibvec}{\vec{\fib}}       
\+{\fnb}{\beta}               
\+{\fng}{\gamma}              
\+{\fne}{\epsilon}            
\+{\g}{\mx{g}}                
\+{\Ie}{I_{\mathrm{ext}}}     
\+{\Ieff}{I_{\mathrm{eff}}}   
\+{\Iion}{I_{\mathrm{ion}}}   
\+{\Iu}{I_u}                  
\+{\Iv}{I_v}                  
\+{\i}{i}                     
\+{\j}{j}                     
\+{\k}{k}                     
\+{\L}{\mathcal{L}}           
\+{\Len}{L}                   
\+{\Linf}{L^\infty}           
\+{\Ltwo}{L^2}                
\+{\n}{n}                     
\+{\normv}{\vec{n}}           
\+{\norm}{n}                  
\+{\r}{{\vec r}}              
\+{\sig}{\sigma}              
\+{\sigi}{\hat{\sig}_{\mathrm{i}}} 
\+{\sige}{\hat{\sig}_{\mathrm{e}}} 
\+{\sigeff}{\hat{\sig}_{\mathrm{eff}}} 
\+{\svr}{\chi}                
\+{\T}{T}                     
\+{\t}{t}                     
\+{\u}{u}                     
\+{\V}{V}                     
\+{\Va}{V^*}                  
\+{\v}{v}                     
\+{\weight}{W}                
\+{\p}{\mx{p}}                
\+{\phie}{\Phi_e}             
\+{\phii}{\Phi_i}             
\+{\phiia}{\Phi^*_i}          
\+{\pin}{_{\textrm{pin}}} 
\+{\q}{\mx{q}}                
\+{\x}{x}                     
\+{\xc}{x_*}                
\+{\y}{y}                     
\+{\yc}{y_*}                
\+{\z}{z}                     
\+{\A}{A}                     
\+{\B}{B}                     
\+{\mS}{\mx{S}}               
\+{\mD}{\mx{\Lambda}}         
\+{\mT}{\mx{T}}               
\usepackage{color}
\newcommand{\del}[1]{}
\renewcommand{\u}{\mx{u}}               
\renewcommand{\D}{\mx{D}}                 
\newcommand{\Q}{\mx{Q}}                 
\newcommand{\paralp}{\alpha}            
\newcommand{\parbet}{\beta}             
\newcommand{\pargam}{\gamma}            

\begin{document}
\onecolumn
\firstpage{1}

\title[{Cardiac Re-entry in micro-CT
    models}]{{Dynamics of cardiac re-entry in
      micro-CT and serial histological
sections based models of mammalian hearts}}

\author[\firstAuthorLast ]{\Authors} 
\address{} 
\correspondance{} 

\extraAuth{}

\maketitle

\begin{abstract}

\section{}
Cardiac re-entry regime of self-organised abnormal synchronisation and
transition to chaos underlie dangerous arrhythmias and fatal
fibrillation. Recent advances in the theory of dissipative vortices,
experimental studies, and anatomically realistic computer simulations,
elucidated the role and importance of cardiac re-entry anatomy induced
drift and interaction with fine anatomical features of the heart.
The fact that anatomy of the heart is consistent
  within a species suggested a possible functional effect of the heart
  anatomy and structural anisotropy on
  spontaneous drift of cardiac
  re-entry. A comparative
study insight into the anatomy induced drift could be used \emph{e.g.} to predict, 
given a specific atrial anatomy, atrial arrhythmia evolution, in order to ultimately improve low-voltage defibrillation protocols and/or ablation strategies.
        
In this paper, in micro-CT based model of
rat pulmonary vein wall, and
    in sheep atria models based on high resolution serial histological
sections, we demonstrate effects of heart geometry and
anisotropy on cardiac re-entry anatomy induced drift, and pinning to
sharp fluctuations of thickness in the tissue layer. The data sets of  sheep
atria and rat pulmonary vein
wall are incorporated into
the BeatBox High Performance Computing simulation environment for
anatomically realistic computer simulations. Cardiac
re-entry is initiated at prescribed locations in the spatially homogeneous
mono-domain micro-CT and serial histological
sections based models of cardiac tissue. Excitation is described by simplified FitzHugh-Nagumo
kinetics. %
In the in-silico models,
isotropic and anisotropic conduction show specific anatomy effects and
the interplay between anatomy and anisotropy of the heart. 
The main objectives are to demonstrate the functional role of the
species hearts geometry and anisotropy in the anatomy induced drift of
cardiac re-entry. In case of the particular region of the rat pulmonary
vein wall with $\sim 90 \degree$ transmural fiber rotation, it is
shown that the joint effect of the PV wall geometry and anisotropy can turn a plane
excitation wave into a re-entry pinned to a small fluctuation of
thickness in the wall. 

\tiny
 \keyFont{ \section{Keywords:} cardiac arrhythmias, anatomically realistic modelling,
   anisotropy, anatomy induced drift, FitzHugh-Nagumo model} 
\end{abstract}

\section{Introduction}

Cardiac re-entry regime of self-organisation and transition to chaos
underly dangerous arrhythmias and fatal
fibrillation~\citep{Mines-1913, Garey-1914, Allessie-etal-1973,
  Pertsov-etal-1993, Christoph-etal-Nature_2018}. 
Tissue and whole organ experiments showed that cardiac
arrhythmias are caused by a combination of
electrophysiological~\citep{BoschNattel-CardiovascularResearch2002,Workman-etal-HeartRythm2008,Kushiyama-etal-2016}, structural~\citep{Pellman-etal-2010,Eckstein-etal-2011,Takemoto-etal-2012,Eckstein-etal-2013},
and anatomical~\citep{MacEdo-etal-2010,Anselmino-etal-2011} factors,
which sustain cardiac
re-entry~\citep{GrayPertsovJalife-1996-Circ,Wu-etal-1998-CR,Nattel-Nature2002,Yamazaki-etal-2012-CVR}. 
That is why, for many decades, cardiac re-entry,
  its origin and a possibility of smooth control and defibrillation,
  have seen extensive mathematical study and computer modelling~\citep{Wiener-Rosenblueth-1946,Balakhovsky-1965,Krinsky-1968,Panfilov-etal-1984,Davydov-etal-1988,Keener-1988,Ermakova-etal-1989,Biktashev-Holden-1994,awt,Fenton-Karma-1998,Pertsov-etal-PRL2000,Wellner-etal-PNAS2002,swd,orbit,Biktashev-etal-2011-PONE,Biktasheva-etal-2015-PRL,
Biktasheva-etal-HF_2018}.

Cardiac
myocyte orientation mapping in mammalian hearts showed that transmural fiber
arrangement is consistent within a
species, and varied between species, including the range of transmural change of fiber angle in
ventricular wall~\citep[p.~173]{Hunter-etal-CompBiolOfHeart}, which therefore
could have a functional role in dynamics and evolutionary control of cardiac
arrhythmias~\citep{Bishop-etal-2010,Bishop-etal-2011,Bishop-Plank-2012,Fukumoto-etal-2016},
also studied in simplified mathematical and computer
models~\citep{Fenton-Karma-1998,Pertsov-etal-PRL2000,Wellner-etal-PNAS2002,  Spach-CircRes-2001, Smaill-etal-2004,
RodriguezEasonTrayanova-2006,Dierckx-etal-PRE2013}. On the
other hand, the recent discovery of the new phenomenon of dissipative
  vortices interaction with small variations of thickness in the
  layer~\citep{Biktasheva-etal-2015-PRL, Ke-etal-Chaos_2015} clarified
  the possible effect of fine anatomical structures in the heart,
  such as \emph{e.g.} Pectinate Muscles (PM), on the anatomy induced
  drift of cardiac re-entry~\citep{Wu-etal-1998-CR,Yamazaki-etal-2012-CVR,Kharche-etal-BioMedRI_2015}.

With the recent advance in High Performance Computing (HPC), it has
become possible for high resolution DT-MRI, serial histological sections, and micro-CT
data sets, containing both detail heart anatomy and myofiber
structure, to be directly incorporated into the computationally
demanding complete anatomy
realistic computer simulations~\citep{Vigmond-etal-ExperimentalPhysiology-2009,
  Bishop-etal-2010, Kharche-etal-FIMH_2015,
  Kharche-etal-BioMedRI_2015, bbx-2017-PONE, Biktasheva-etal-HF_2018}, 
in order to see in the \emph{in-silico} models the specific anatomy
effects on cardiac re-entry, as well as the role of the interplay
between the anatomy of the heart and its structured anisotropy.

In this paper, we present an anatomy and myofiber structure  
realistic computer simulation study of cardiac re-entry dynamics in sheep atria models based on serial histological sections data sets, 
and in rat pulmonary vein wall model based on micro-CT data sets. %
The main objectives are to demonstrate: i) the
  functional role of geometry and structured anisotropy of the heart for the anatomy induced drift of
cardiac re-entry;  ii) that the anatomical settings
of the isolated sheep atria, and that of a rat pulmonary vein wall, might support
a positive filament tension re-entry; iii) for the particular region of rat pulmonary
vein wall with $\sim 90 \degree$ transmural fiber rotation, it is
shown that the combination of geometry and anisotropy can turn a plane
excitation wave into a cardiac re-entry pinned to a small fluctuation of the wall thickness. 

We demonstrate that, in the sheep atria, positive filament tension re-entry might pin to
Bachman’s bundle and to pectinate muscles, which would  
sustain the re-entry. 
In rat pulmonary vein wall, where the transmural fiber direction 
change of $\sim90\degree$ seems to be a salient feature of the wall
anatomy, an otherwise stable isotropic 
micro anatomic reentry might produce qualitatively different endocardial and
epicardial manifestation, so that the
micro anatomic reentry clearly seen on the endocardium might appear as a
focal point on the epicardium.  The same region of the rat pulmonary
vein with $\sim 90 \degree$ transmural fiber rotation can turn a plane
excitation wave into a cardiac re-entry pinned to a nearby small fluctuation of the wall thickness.

The main significance of our initial findings is
that the computationally
demanding high resolution serial histological sections and micro-CT
data sets can be directly incorporated into the complete anatomy and anisotropy realistic HPC computer
simulations, and thus serve a model for intra- and inter-species comparative in-silico study of cardiac re-entry dynamics in mammalian hearts.

\section{Materials and Methods}

\subsection{Anatomy Data Sets}

\label{Methods}

\subsubsection{Sheep Atrial Serial Section Histology Data Set}
\label{Sheep-DataSet}

\begin{figure}[tb] \centering
\includegraphics[width=\linewidth]{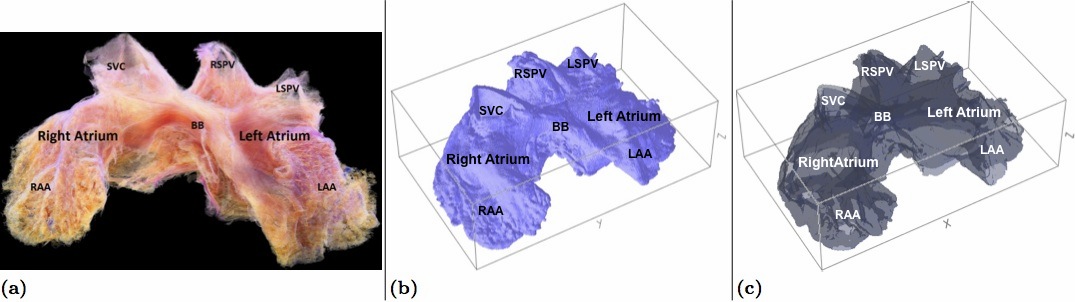}%
\caption{\label{Fig1} {\bf Sheep atrial
    model}. Translucent  views of sheep atria displayed from an
  anterior angle. (a-b) Atrial geometry model: grid size $597 \times 935 \times 327$,
  resolution $0.1mm$.(c) Down-sampled atrial geometry model: grid size $313 \times
  200 \times 111$,
  resolution $0.3mm$. 
}
\end{figure}

Atria from a crossbred sheep were extracted and processed. The tissue
was fixed in formalin, and a graded series of ethanol was used for
dehydration. Then the specimen was embedded in paraffin wax. Tissue
architecture throughout the atria was reconstructed using the extended
volume surface imaging modality. The images were acquired at a
resolution of $8.33 \mu m^2$ per pixel after staining the tissue
surface using Toludine blue. After acquiring each image, the upper
surface of tissue was milled off with an ultramiller. Atrial surface
images from top to bottom were acquired at every $50 \mu m$ steps by
following this approach of milling, staining and imaging. The
resulting tissue volume was segmented using a custom-designed suite of
image processing tools. The segmented 2D sections were down-sampled to
produce an isotropic volume of $100\times100\times100 \mu m^3$ voxels. Fiber orientation field over this geometry was defined using structure tensor approach. More detailed accounts on tissue processing, image acquisition, image processing and fibre field computation steps can be found elsewhere \citep{Zhao2012, Gerneke2007}.

\subsubsection{Sheep Atrial Serial Section Histology based anatomy model}

Figure~\ref{Fig1}(a) %
 shows the sheep atrial serial section histology based model of the $597 \times 935 \times 327$ voxels size, with  resolution of
$100\mu m$.  The serial section histology model was converted into the
\bbx~\citep{bbx-2017-PONE} regular Cartesian mesh \code{.bbg} geometry
format, also shown in Figure~\ref{Fig1}(b), %
containing the coordinates of the atria
tissue points together with the local fibre
orientation~\citep{bbx-2017-PONE}. Following~\citep{bbx-2017-PONE},
  a \bbg\ file is a \code{csv}  ASCII 
file describing tissue points in the regular mesh, one tissue point per line in the format: %
\[
 \code{ x,y,z,status,fibre\_x,fibre\_y,fibre\_z}
\]
where \code{x, y, z} are integer Cartesian coordinates of a voxel,
\code{status} is a flag with a nonzero-value for a tissue point, and \code{fibre\_x, fibre\_y, fibre\_z} are $x$-,
$y$- and $z$-components of the fibre orientation vector at that
point. Only tissue points with nonzero \code{status} need to be specified. The
minimal bounding box for sheep
atria model is
$597 \times 935 \times 327$ voxels size, and contains 25,124,362
tissue (non-void)
points. 

 From the original sheep atrial serial section histology model shown
 in Figure~\ref{Fig1}(a), we also made the sheep atria anatomy model
\emph{down-sampled} three times in each of the
    \code{x,y,z} dimensions, contained in a box of
    $200 \times 313 \times 111$ voxels, resolution $0.3mm$, shown in
    Figure~\ref{Fig1}(c). The \emph{down-sampled} sheep atria model contained 932,374 tissue points.

\subsubsection{\label{PVWallDataSet}Rat Pulmonary Vein Wall micro-CT Data Set}
\label{PVWall-microCT}

\begin{figure}[tb] \centering
\includegraphics[width=\linewidth]{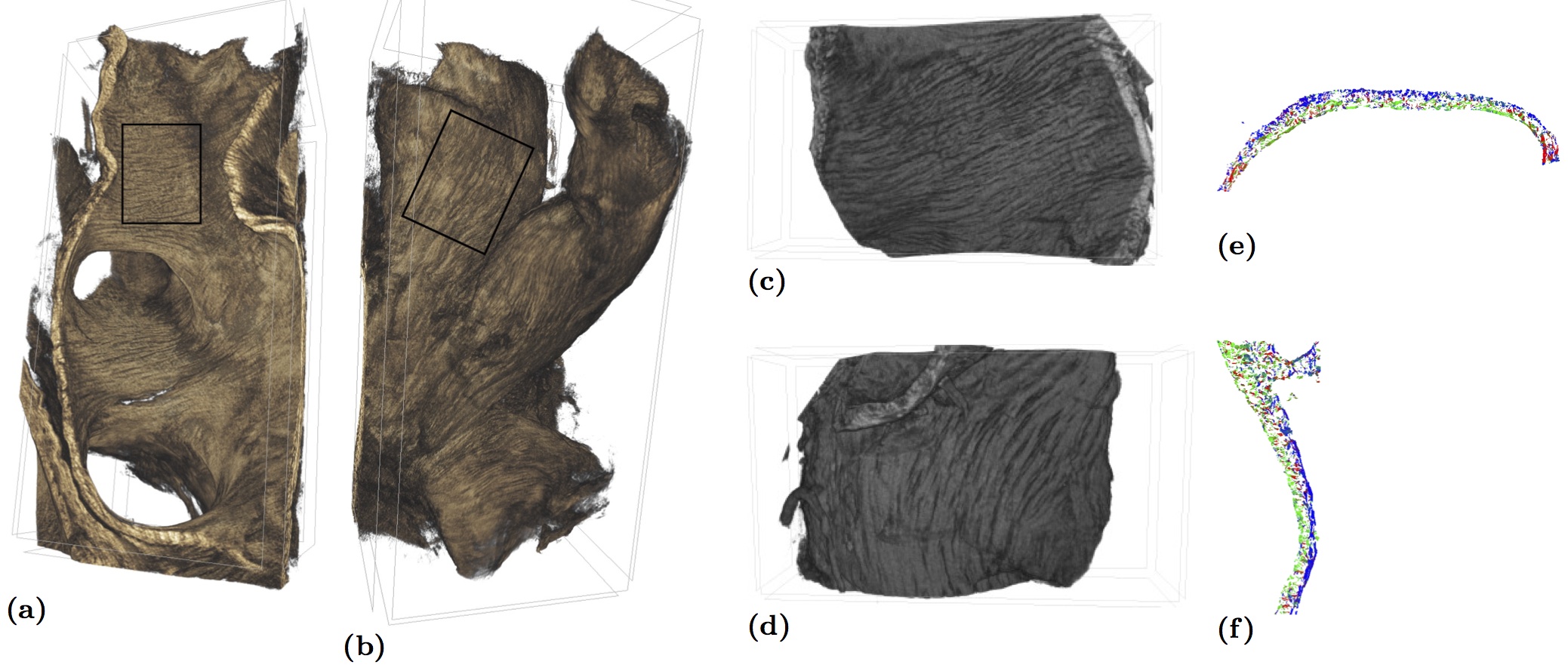}%
\caption{\label{Fig2} {\bf Rat Pulmonary Veins and  cropped PV
  wall micro-CT}: {\bf (a)} Coronal view of
  the endocardium of the PVs, {\bf (b)} Epicardial view of PVs from the
  superoposterior view. The rectangles in the panels (a-b) show the area corresponding to
  {\bf (c)} the endocardial, and {\bf (d)} epicardial views of the cropped PV
  wall micro-CT, resolution $3.5\mu m$. See
  also Supplementary Material movie Fig2cd-rev.mpg. {\bf (e)}  color coded fibre
  orientation in the horisontal middle PV wall cross section, and {\bf
    (f)}  fibre orientation in the vertical middle PV wall cross section:
  \code{red} represents \code{z} component of the local fiber orientation
  vector, \code{green} represents \code{x} component of the local fiber orientation
  vector,  and \code{blue} represents \code{y} component of the local fiber orientation
  vector. It can be seen that there is predominantely
  \code{blue} that is along the \code{y} axis fiber orientation at the
  epicardum vs predominantely \code{green} that is along \code{x} axis fiber
  orientation at the endocardium of the rat PV Wall.
}
\end{figure}

Hearts from Wistar rats were excised after thoracotomy of the anaesthetised animal. The aorta was cannulated and flushed with ice cold saline solution. The pulmonary veins and vena cavae were ligated and liquid agar was injected into the chambers to keep the atria in an ``inflated'' state. The agar was allowed to set and the tissue was fixed in 4\% paraformaldehyde for 48 hours at 4\degree C. After fixation, excess fat and connective tissue was dissected from the epicardium. And the atria were then transected at the atrioventricular annulus and isolated from the ventricles.

The tissue was then dehydrated by immersion in aqueous ethanol in a graded series (30\% to 70\% at 10\% intervals) and stained for 48 hours with a solution of 3\% phosphomolybdic acid in 70\% aqueous ethanol. The agar was removed from the tissue, which was then washed and blotted to remove excess stain. The sample was fitted into a polypropylene tube and secured using pieces of polystyrene cut to fit the tube and mounted onto the stage of the micro-ct (Skyscan 1272, Bruker, Kontich, Belgium). Single images were initially taken and compared at 15-minute intervals to assess sample shrinkage due to dehydration. After the stabilisation period, imaging parameters were optimised to obtain satisfactory tissue-air contrast. The resulting voxel resolution was $3.5\mu m^3$, and $360\degree$ scans were performed at $0.2\degree$ intervals. A $0.25mm$ Aluminium filter was used to block low energy x-rays during the scan.

3D Reconstruction ( Figure~\ref{Fig2}(a-d)) was performed using the provided software (InstaRecon, Bruker, Kontich, Belgium) using the recommended settings. The images were then processed and cropped using a custom made GUI based on ITK \citep{itk} and VTK \citep{vtk} libraries written in \code{c++}. Images were ``resliced'' in 3D in order to align the axes with the orientation of the fibres. The structure tensor analysis algorithm \citep{Zhao2012} was used to assign a 3D direction to each voxel representing a fibre.

\subsubsection{Rat Pulmonary Vein Wall micro-CT based model}

Figure~\ref{Fig2}(c,d) %
 shows the endocardial and epicardial views of the rat pulmonary vein
 wall micro-CT model of the $254 \times 814 \times 543$ voxels size, with  resolution of
$3.5\mu m$.  The  micro-CT model was converted into the
\bbx~\citep{bbx-2017-PONE} regular Cartesian mesh \code{.bbg} geometry
format, %
containing the coordinates of the PV wall
tissue points together with the local fibre
orientation~\citep{bbx-2017-PONE}. The  minimal bounding box of the
rat PV wall anatomy model is
$254 \times 814 \times 543$ grid size, and containes 230,355,342  tissue
points. 

Figure~\ref{Fig2}(e,f) %
 panels show color coded fibre orientation in the horisontal and
 veritical middle cross sections of the rat PV wall. Here, the \code{rgb} color
 coding scheme was chosen so that \code{red} represents \code{z} component of the local fiber orientation
  vector, \code{green} represents \code{x} component of the local fiber orientation
  vector,  and \code{blue} represents \code{y} component of the local fiber orientation
  vector.  In full accordance with the original images
  fibre orientation alignment along the axes, as described in the
  previous section~\ref{PVWall-microCT}, it can be seen in
  Figure~\ref{Fig2}(e,f) that the fiber orientation at the
  epicardum side of the wall is predominantly
  \code{blue}, that is aligned along the \code{y} axis, while the endocardium fiber
  orientation of the rat PV wall is predominantely \code{green}, that is the fibres at the inner side of the PV wall are aligned
  along the \code{x} axis, that is there is $\sim 90\degree$
  transmural change in fibre orientation in the rat PV wall.

\subsection{Tissue model}
\label{RD}

Similar to our previous
  paper~\citep{Biktasheva-etal-HF_2018}, all the three (``complete''  and ``down-sampled'' sheep atria shown in Figure~\ref{Fig1}(b,c), and the rat PV
wall shown in Figure~\ref{Fig2}(c,d)) data sets were used to
implement corresponding \textit{monodomain} tissue models with non-flux boundary conditions 
\begin{align} &
  \df{\u}{\t} = \f (\u)
  + \nabla\cdot\hat\D\nabla \u ,  
			    \label{bc} \\ &
  \qquad \qquad {\normv \cdot\hat\D\nabla \u}\bigg|_G = 0.  
			    \nonumber 
\end{align}
Here  $\u(\r,t)={(u, v)}^T$, $\r$ is the position vector, $\f(\r,t)={(f, g)}^T$	
is the FitzHugh-Nagumo~\citep{Winfree-1991} kinetics column-vector 
\begin{align}
  f(u,v) &= \paralp^{-1}(u-u^3/3-v),                         \nonumber\\
  g(u,v) &= \paralp \, (  u + \parbet - \pargam v ), \label{FHN}
\end{align}
 with the kinetics parameters $\paralp=0.3$, $\parbet=0.71$,  $\pargam=0.5$,
   which in perfectly homogeneous infinite medium correspond
 to a rigidly rotating vortex with positive filament
 tension~\citep{ft}. The specific forces generated by
 anisotropy~\citep{Wellner-etal-PNAS2002, Dierckx-etal-2015-PRL, Biktasheva-etal-HF_2018} and by small
 fuctuations of thickness in the layer~\citep{Biktasheva-etal-2015-PRL, Kharche-etal-BioMedRI_2015} are only two types out of a variety of specific forces acting on cardiac re-entry in the heart~\citep{Biktashev-etal-2011-PONE}. For this study, we have intentionally chosen the simplified FHN model with the
 parameters corresponding to the simplest possible rigidly rotating
 positive filament tension re-entry, in order to eliminate effects of
 \emph{e.g.} meander~\citep{Winfree-1991},
 alternans~\citep{Karma_Chaos1994}, negative filament
 tension~\citep{ft},  etc., and enhance and highlight pure effects of
 the tissue geometry and anisotropy on the cardiac re-entry dynamics. The diffusion matrix tensor is defined as
 $\hat\D=\Q\hat P$, where $\Q=\mathrm{diag}(1,0)=\begin{bmatrix}1&0\\0&0\end{bmatrix}$ is the matrix of the relative diffusion coefficients for $u$ and $v$ components, and $\hat
P=[P_{\j\k}]\in\Real^{3\times 3}$ is the $u$ component diffusion tensor, which has only two different eigenvalues: the bigger, simple eigenvalue $\Ppar$ corresponding to the direction along the tissue fibers, and the smaller, double eigenvalue $\Port$, corresponding to the directions across the fibres, so that
\begin{equation}
  P_{\j\k} = \Port\kron_{\j\k} +
  \left(\Ppar-\Port\right)\fib_\j\fib_\k,
                                        \label{sigma}
\end{equation}
where $\fibvec=\left(\fib_\k\right)$ is the unit vector of the fibre
direction; $\normv$ is the vector normal to the tissue boundary
$G$. In the isotropic
simulation, $\Ppar$ and $\Port$ values were fixed at
$\Ppar=\Port=1$ (corresponding 1D conduction velocity 1.89, in the dimensionless
  units of Eqs.~(\ref{bc})-(\ref{FHN})). In the anisotropic simulations, $\Ppar$ and $\Port$
values were fixed at $\Ppar=2$, $\Port=0.5$ (corresponding conduction
velocities 2.68 and 1.34 respectively). All the conduction velocities
have been computed for plane periodic waves with the frequency of the free spiral wave in the model, i.e. 11.36. With the isotropic
diffusivity ($\Ppar=\Port=1$) equal to the geometric mean between the faster and the slower anisotropic diffusivities ($\Ppar=2, \Port=1/2$), the isotropic conduction velocity 1.89 was almost exactly the same as the geometric mean $\approx1.895$ of the
faster and slower (2.68 and 1.34 respectively) anisotropic conduction
velocities, chosen in order to minimize the maximal relative
difference between the isotropic and anisotropic propagation speeds.

All the computer simulations presented here were done using the \bbx~\citep{bbx-2017-PONE} 
software package with the explicit time-step Euler
scheme on Cartesian regular grid, 7-point stencil Laplacian
approximation for \emph{isotropic simulations}, 27-point stencil for
\emph{anisotropic} Laplacian approximation.
In most simulations, unless specified otherwise, re-entry was initiated by the phase distribution
method~\citep{chaos} at prescribed locations of the respected three (``complete'' and ``down
sampled'' sheep atria, and the rat PV wall) anatomy models described above.

The space and time discretization
    steps listed in Table~\ref{tab:steps} were chosen for each tissue
    model in order to ensure computational stability of the
    \bbx~\citep{bbx-2017-PONE} explicit time-step forward Euler
    scheme. Both the time step and the space step are given in the
  dimensionless units of equations~\eqref{bc}--\eqref{FHN}. Note that the
    ``physiological'' space scale is recovered by noting that $\Delta x$
    corresponds to the voxel size of the corresponding grid as described in
    Section \ref{Methods}.
  \begin{table}[tb] 
    \centering
    \begin{tabular}{|l|l|l|l|l|} 
\hline 
      \rule[-1ex]{0pt}{3ex} {\bf Geometry} & {\bf \emph{(an)}isotropy} & {\bf Figures} & $\Delta x$ & $\Delta t$ \\[3pt]\hline
      \rule[-1ex]{0pt}{3ex} Sheep atria ``down-sampled'' &
      isotropic &
      \ref{Fig1}(c), \ref{Fig3}(a--c) &
      $0.15$ & $0.003$ \\[3pt]\hline
      \rule[-1ex]{0pt}{3ex} Sheep atria ``complete'' &
      isotropic &
      \ref{Fig1}(b), \ref{Fig4}, \ref{Fig5}, \ref{Fig6}(a,b) &
      $0.1$ & $0.001$ \\[3pt]\hline
      \rule[-1ex]{0pt}{3ex} Sheep atria ``complete'' &
      anisotropic &
      \ref{Fig1}(b), \ref{Fig3}(d), \ref{Fig4}, \ref{Fig5}, \ref{Fig6}(a,c) &
      $0.1$ & $0.0005$ \\[3pt]\hline
      \rule[-1ex]{0pt}{3ex} Rat PV wall &
      isotropic &
      \ref{Fig2}(c,d), \ref{Fig7}--\ref{Fig10} &
      $0.1$ & $0.001$ \\[3pt]\hline
      \rule[-1ex]{0pt}{3ex} Rat PV wall &
      anisotropic &
      \ref{Fig2}(c,d), \ref{Fig11} &
      $0.1$ & $0.0005$ \\[3pt]\hline
    \end{tabular}
    \caption[]{\label{tab:steps} Discretization parameters used in simulations: space step discretization $\Delta x$, time step discretisation $\Delta t$.}
  \end{table}






\vspace{5pt}

\section{Results}
\label{RESULTS}

\subsection{\label{Sheep_results} Sheep Atria model}
\subsubsection{\label{Down-Sampled} Down-Sampled Isotropic Sheep Atria model}
\begin{figure*}[tb]
\includegraphics{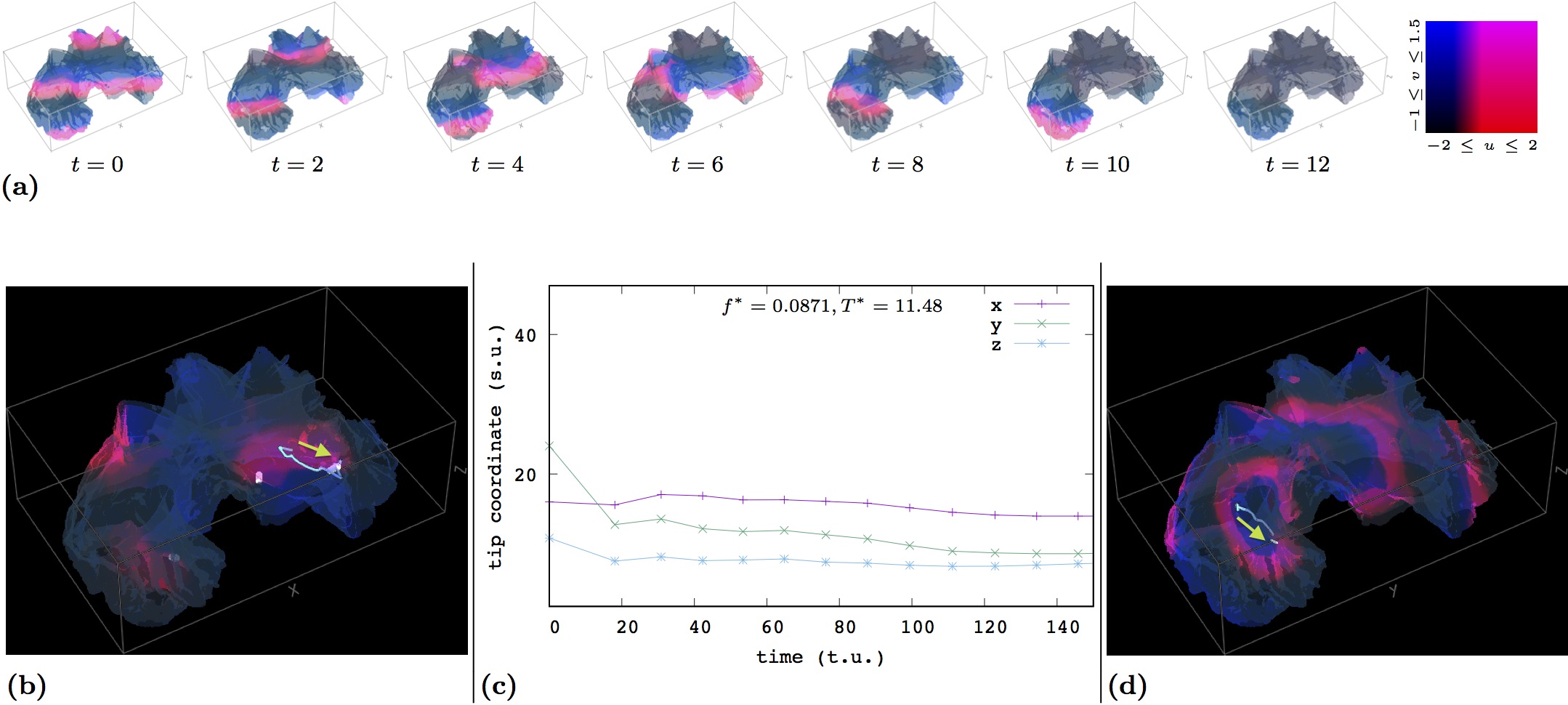}%
\caption{\label{Fig3} 
{\bf Sheep Atria simulations.} 
 {\bf (a) Down-sampled Isotropic model: plane excitation wave}. 
 The translucent  Sheep  Atria is shown in grey, excitation front
 shown in red, time shown under each panel in time units of Eqs.~(\ref{bc})-(\ref{FHN}). The excitation
   wave initiated just below Pulmonary Veins propagates through the
   isotropic atria and terminates in the RAA. 
 {\bf (b) Down-sampled Isotropic model: Re-entry pinning in the LAA}:
 Re-entry is initiated in the Left Atrium. Filaments of the initial
 and of the final
 pinned re-entry are shown in white; drift trajectory is shown as thin
 blue line; green arrow shows direction of the drift. Filament of the secondary anatomical (not drifting) re-entry is also seen close to BB.
{\bf (c) Down-sampled Isotropic model}: time course in time units of Eqs.~(\ref{bc})-(\ref{FHN}) of the coordinates
of the filament's epicardium end of the drifting
re-entry shown in the previous panel (b). The dominant frequency and
the dominant period of the drifting re-entry are shown
   at the top of the panel.
{\bf (d) ``Complete'' Anisotropic model: Re-entry pinning in the RAA}:
 Re-entry is initiated in the  Right Atrium. 
Filaments of the initial
 and of the final
 pinned re-entry are shown in white; drift trajectory is shown as thin
 blue line; green arrow shows direction of the drift. 
Convergence (time course) of the coordinates of the epicardium end of the drifting filament to
   its final ``pinned'' location, together with the corresponding
   dominant frequency and dominant
   period of the drifting re-entry are shown in Figure~\ref{Fig5}(h).
 }
\end{figure*}

\begin{figure*}[tb]
\includegraphics{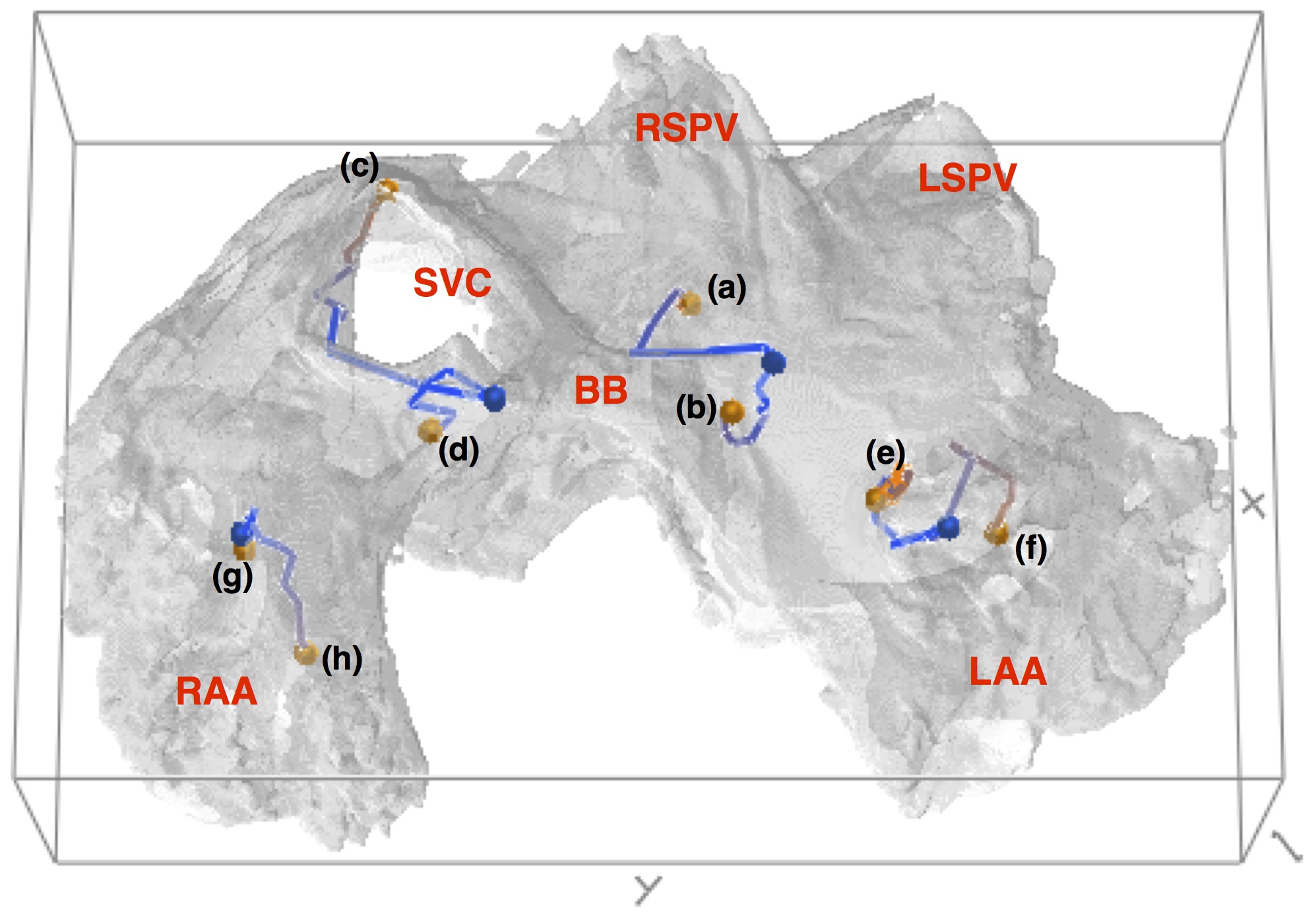}%
\caption{\label{Fig4}  
{\bf ``Complete'' Sheep Atria model: trajectories of Isotropic vs
  Anisotropic drift}. The translucent  Sheep  Atria
(Figure~\ref{Fig1}(b)) is shown in grey. Pairs of isotropic and anisotropic drift trajectories start from the same initial
locations shown as blue dots. The final pinning locations are shown as orange dots. Drift trajectories are shown as lines
of changing color from blue in the beginning of a trajectory to
orange in the end of a trajectory. Letter in the brackets next to the end of each trajectory corresponds to the
panel label in Figure~\ref{Fig5} showing the time course of the coordinates of the epicardium end of the drifting filament to
   its final ``pinned'' location. See also (Multimedia view) Fig4-rev.mpg.
}
\end{figure*}

\begin{figure*}[tb]
\includegraphics[width=\textwidth]{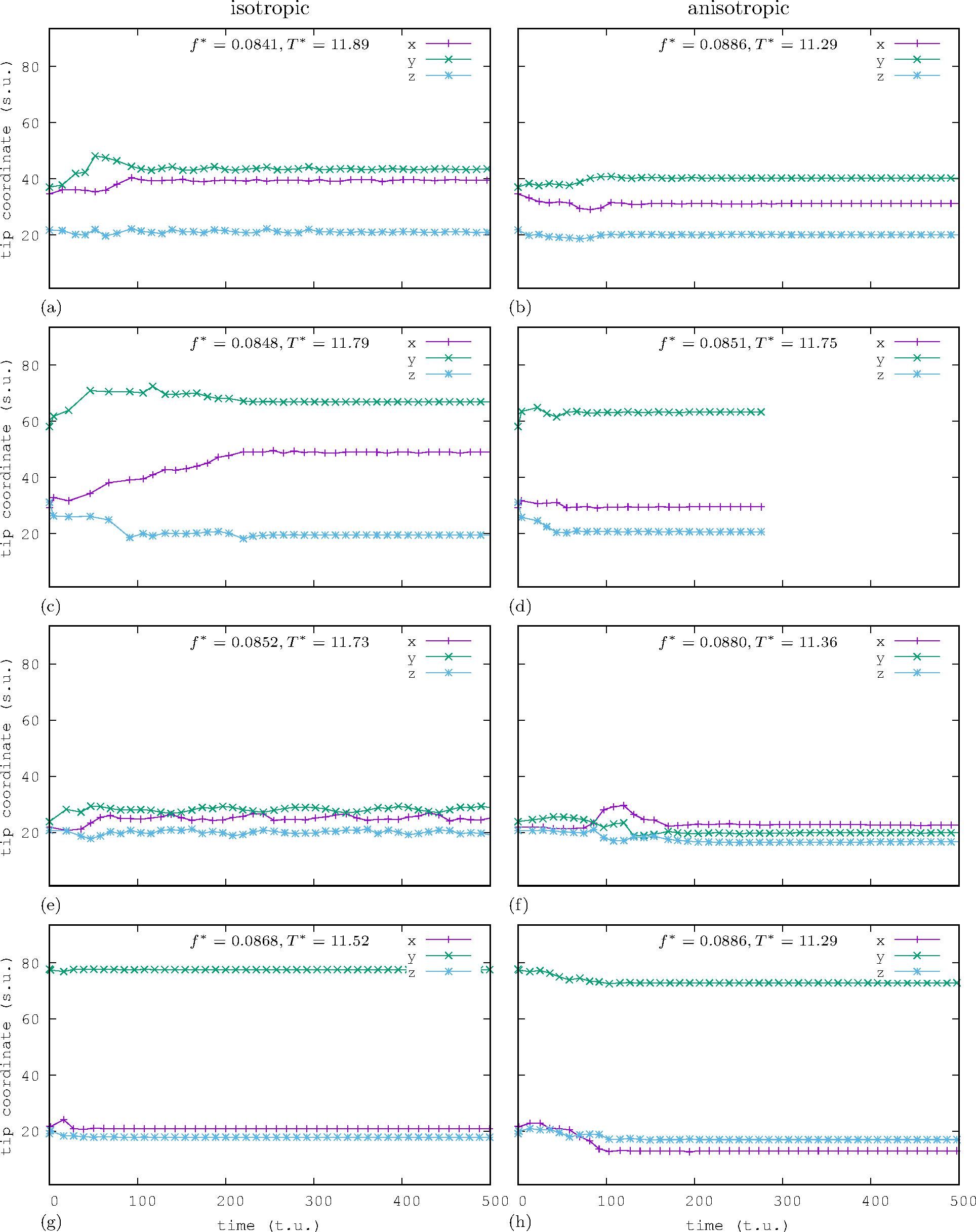}%
\caption{\label{Fig5} 
{\bf ``Complete'' Sheep Atria model: Isotropic vs
  Anisotropic drift}. Each panel (a-h) shows time course
of the corresponding trajectories shown in Figure~\ref{Fig4}, together with the dominant frequency and
   period of the corresponding drifting re-entries. 
}
\end{figure*}

\begin{figure*}[tb]
\includegraphics[width=\textwidth]{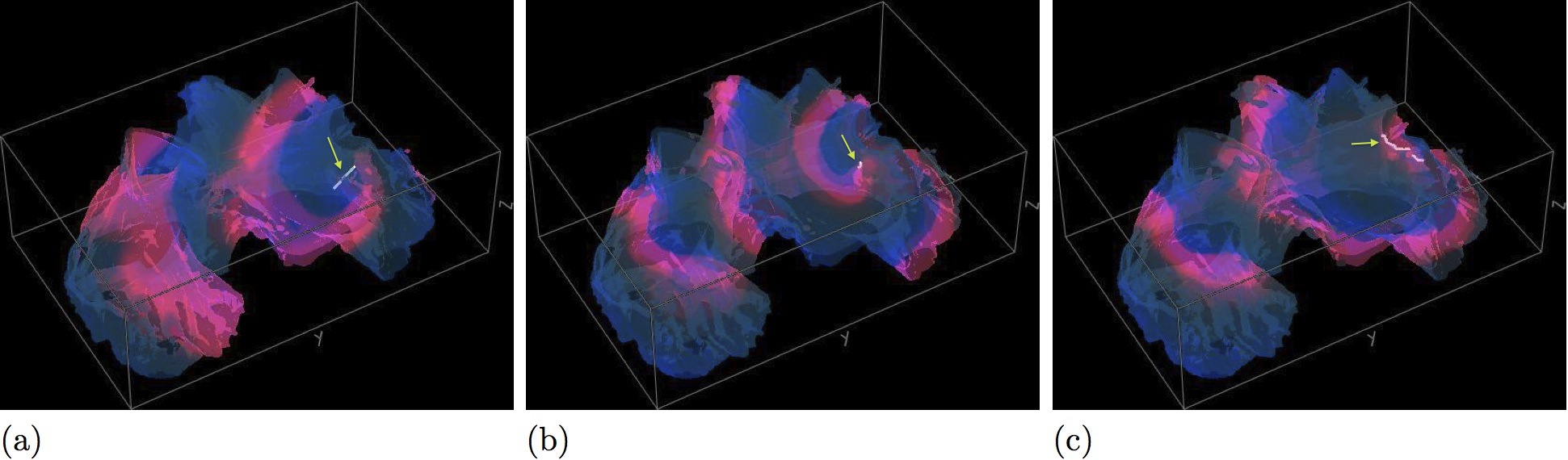}%
\caption{\label{Fig6} 
{\bf ``Complete'' Sheep Atria model:  Different lengths of re-entry
  filaments in Isotropic vs
  Anisotropic conductance}.  The translucent  Sheep  Atria
(Figure~\ref{Fig1}(b)) is shown in grey, excitation front shown in red (see the color box in
  Figure~\ref{Fig3}(a)). Green arrows point to the white instant
  filaments. 
{\bf (a)} \emph{Initial transmural} filament of the pair of isotropic
``(e)'' and anisotropic ``(f)'' trajectories shown in  Figure~\ref{Fig4} ;
{\bf (b)} after 11 rotations, very short \emph{transmural} filament in the
\emph{isotropic} trajectory ``(e)'' shown in  Figure~\ref{Fig4} ; 
{\bf (c)}  after 11 rotations, long \emph{intramural} filament in the
\emph{anisotropic} trajectory ``(f)'' shown in  Figure~\ref{Fig4}. 
The corresponding time courses
  of the coordinates of the epicardal ends of the filaments are shown in
   Figure~\ref{Fig5}(e) (iso) and Figure~\ref{Fig5}(f) (aniso).
}
\end{figure*}

For the sake of a basic test of whether a plane excitation wave propagates through
the isotropic sheep atria geometry, a plane excitation wave was
initiated in the down-sampled
  isotropic sheep atria model (Figure~\ref{Fig1}(c)), in the left
  atrium just below the
  right (RSPV) and left (LSPV) Superior Pulmonary Veins,
  see Figure~\ref{Fig3}(a), panel $t=0$. The plane wave propagated through the
   isotropic atria, and terminated in the Right Atrial Appendage
   (RAA), as shown in Figure~\ref{Fig3}(a). %

Figure~\ref{Fig3}(b), %
shows the anatomy induced drift of a clockwise excitation
re-entry initiated by the phase distribution
method~\citep{chaos} in the down-sampled
  isotropic sheep atria model shown in Figure~\ref{Fig1}(c), with the initial position of the transmural filament
at the entrance to the Left Atrial Appendage
(LAA). 
In Figure~\ref{Fig3}(b), the re-entry is shown at its final position pinned
to a pectinate muscle fluctuation of thickness in the
LAA. The re-entry's initial and the final transmural filaments are shown in
white; the trajectory
 of the drifting 
 filament is shown as thin blue line; the green arrow shows direction of
 the drift. The filament of a secondary anatomical (not drifting)
 re-entry can also be seen close to the Bachmann bundle (BB).
 Figure~\ref{Fig3}(c) shows the time course of the coordinates of the drifting filament's epicardium end,  
from its initial position to the pinning point. The  dominant
frequency and the dominant period of the drifting re-entry are shown
   at the top of Figure~\ref{Fig3}(c). From the drift trajectory in
   Figure~\ref{Fig3}(b), and the time course of coordinates of the
   epicardial end of the 
   drifting filament shown in Figure~\ref{Fig3}(c), it can be seen that
   the re-entry's initial transient followed the Pectinate Muscles (PM) structures into the LAA,
until the re-entry's final pinning in the LAA.

\subsubsection{\label{H_ASh} ``Complete'' Sheep Atria model}

Figure~\ref{Fig3}(d) shows the anatomy induced drift of re-entry initiated at the entrance to
  the Right Atrium in the “complete” anisotropic sheep atria model shown in
  Figure~\ref{Fig1}(b). In Figure~\ref{Fig3}(d), the re-entry is shown at its final position pinned
to a fluctuation of thickness of a pectinate muscle in the
RAA. The re-entry's initial and the final transmural filaments are shown in
white; the trajectory
 of the drifting 
 filament is shown as thin blue line; the green arrow shows direction of
 the drift. The time course of coordinates of the drifting
  filament's epicardium end is shown in Figure~\ref{Fig5}(h). The
  dominant frequency and period of the drifting re-entry are shown at
  the top of Figure~\ref{Fig5}(h). Note that the anisotropic drift trajectory of the
  re-entry shown here in Figure~\ref{Fig3}(d) is also shown as
  ``trajectory (h)'' in the collective phase portrate of drift
  trajectories in Figure~\ref{Fig4}. 

Figure~\ref{Fig4} shows a set of pairs of ``isotropic'' and
  corresponding ``anisotropic'' drift trajectories. Each pair of simulations started from
  identical initial conditions at
  different locations in the “complete” model of sheep atria shown in
  Figure~\ref{Fig1}(b). In Figure~\ref{Fig4}, the initial locations of
  the re-entries are shown as blue dots, the final pinning locations are shown as orange dots. Drift trajectories are shown as lines
of changing color from blue in the beginning of a trajectory to
orange in the end of a trajectory. Letters in the brackets next to the
each trajectory pinning point in Figure~\ref{Fig4} correspond to the
panels labelling in Figure~\ref{Fig5}, which shows the time
courses of coordinates of the drifting filaments' epicardium ends. The dominant frequencies
  and dominant periods of the drifting re-entries are also shown at the top of
  each corresponding panel in Figure~\ref{Fig5}. The panels in
Figure~\ref{Fig5}  are arranged in the way that the left column panels
show isotropic drift, and the right column panels show anisotropic
drift. Each horisontal pair of panels in Figure~\ref{Fig5} presents a
pair of isotropic and corresponding anisotropic trajectories starting from the same
initial position. 

Figure~\ref{Fig6} shows a particularly interesting case of the re-entry
    initiated in the Right Atrium, as in this case isotropic vs
    anisotropic conductance result not only in different
    drift trajectories ``(e)'' (iso) and ``(f)'' (aniso) shown in
    Figure~\ref{Fig4} and in Figure~\ref{Fig5}(e-f), but also in
    very different lengths and orientation of the final pinned
    filaments shown here in
    Figure~\ref{Fig6}(b) and Figure~\ref{Fig6}(c). Starting from the
    same initial conditions transmural filament shown in Figure~\ref{Fig6}(a), after 11
    rotations in the isotropic ``complete'' model of the sheep atria (Figure~\ref{Fig1}(b)), the filament becomes a very
    short perfectly \emph{transmural} filament shown in
    Figure~\ref{Fig6}(b). While, after the same number of 11
    rotations in the same very thin but ``anisotropy on'' sheep  atria, the filament
    seen here in Figure~\ref{Fig6}(c) becomes \emph{intramural}, i.e. stretches almost parallel to the epicardial surface. Note
        that theory predicts strong effects of the
        anisotropy on the dynamics of singular filaments,
        e.g. ~\citep{Wellner-etal-PNAS2002}, which must act in
        addition to purely geometric forces. The 
    dominant frequencies and the dominant periods of the pair of isotropic and
    anisotropic re-entries are also shown in 
    Figure~\ref{Fig5}(e) (iso) and Figure~\ref{Fig5}(f) (aniso) correspondingly.

\subsection{\label{Rat_PVWal_results} Rat Pulmonary Vein Wall model}

\begin{figure*}[tb]
\includegraphics{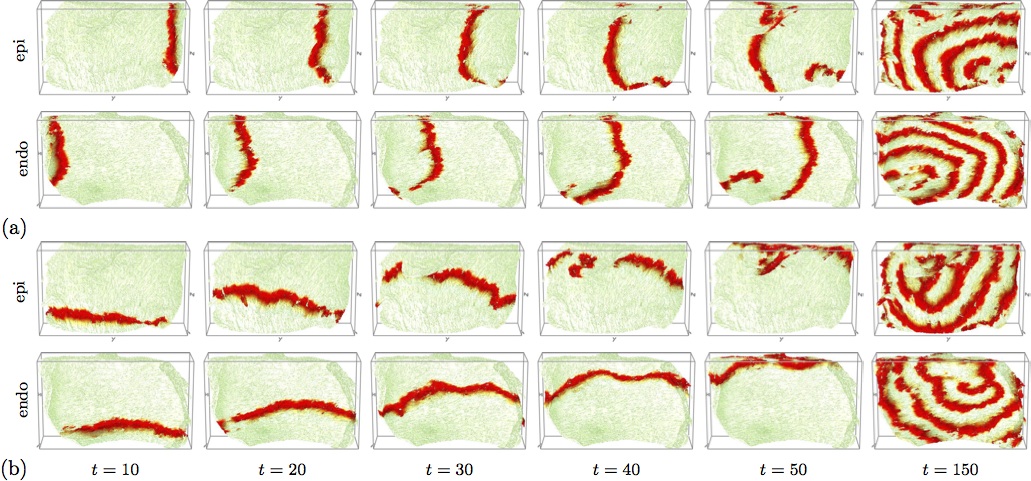}%
\caption{\label{Fig7} 
{\bf  Rat PV Wall model: plane wave
    generates a re-entry.} The Rat Pulmonary Vein Wall is shown
  in yellow, excitation front shown in red (see the color box in
  Figure~\ref{Fig8}); time shown under each column of corresponding panels in time units of Eqs.~(\ref{bc})-(\ref{FHN}).
{\bf (a)}: epicardial and endocardial view of the plane
excitation wave propagating \emph{left to right} (in the endocardial
view), resulting in the re-entry pinned to the cusp of thickness in the bottom left conner of the wall (in the endocardial
view,  see also  Figure~\ref{Fig2}(c)). (Multimedia view)
Fig7a-rev.mpg.
{\bf (b)}: epicardial and endocardial view of the plane
excitation wave propagating \emph{upward} the PV wall,
resulting in the re-entry pinned to the ``handle'' in
the top left conner of the wall (in the epicardial
view,  see also  Figure~\ref{Fig2}(d)). (Multimedia view) Fig7b-rev.mpg.
}
\end{figure*}
\begin{figure*}[tb]
\includegraphics{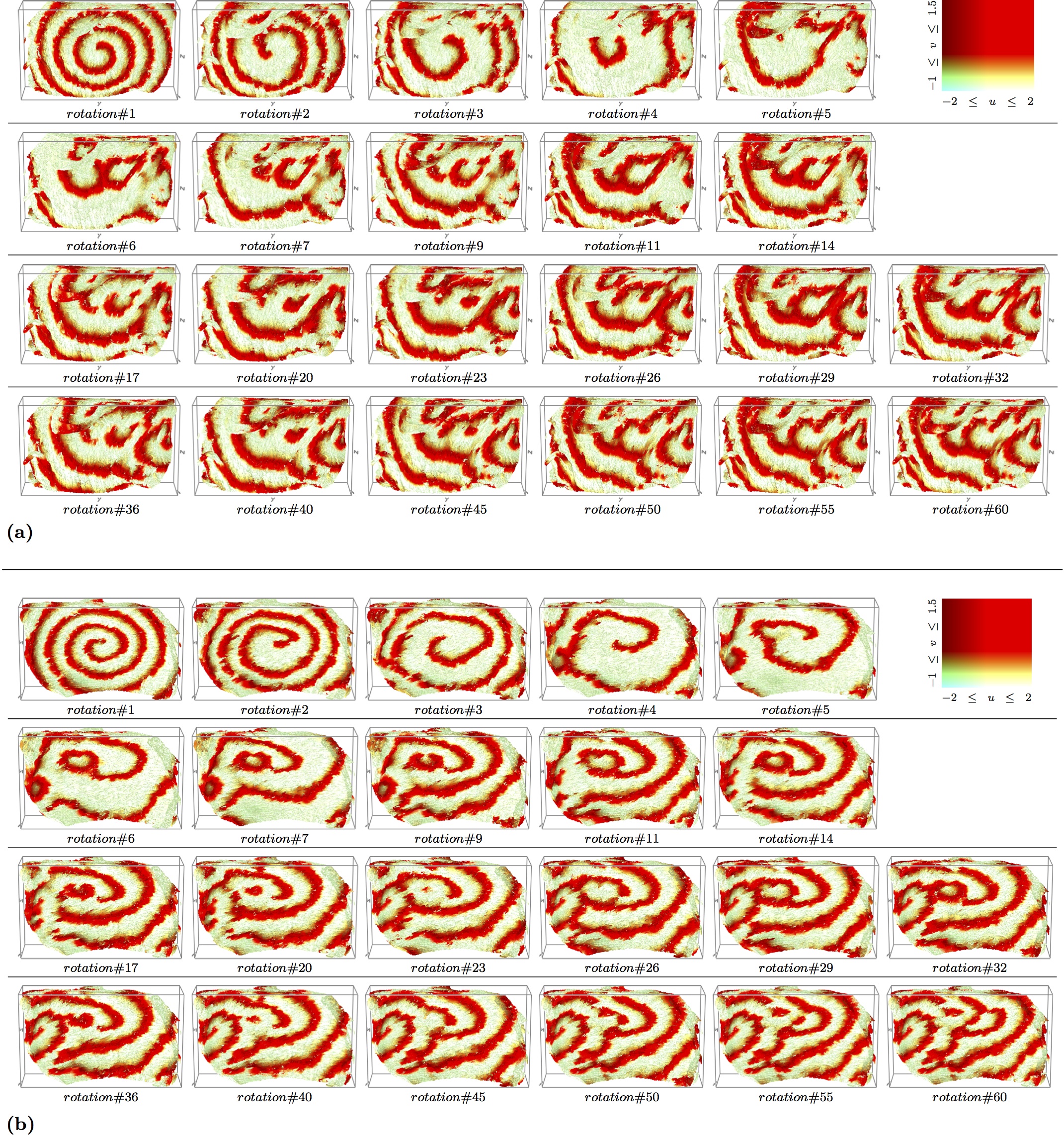}%
\caption{\label{Fig8} {\bf  Rat PV Wall model. Isotropic counter
    clockwise transmural re-entry.} The Rat Pulmonary Vein Wall is shown
  in yellow, excitation front shown in red; stroboscopic visualisation
  of re-entry dynamics: time shown under each
  panel as the re-entry rotation number.
The counter clockwise (in epicardial view) re-entry is initiated transmurally in the middle of the
    PV Wall sample, then fast initial transient is followed by the re-entry
stabilisation. {\bf (a)
  Epicardial view}; (Multimedia view) Fig8a.mpg.
 {\bf (b) Endocardial view}; (Multimedia view) Fig8b.mpg.}
\end{figure*}
\begin{figure*}[tb]
\includegraphics{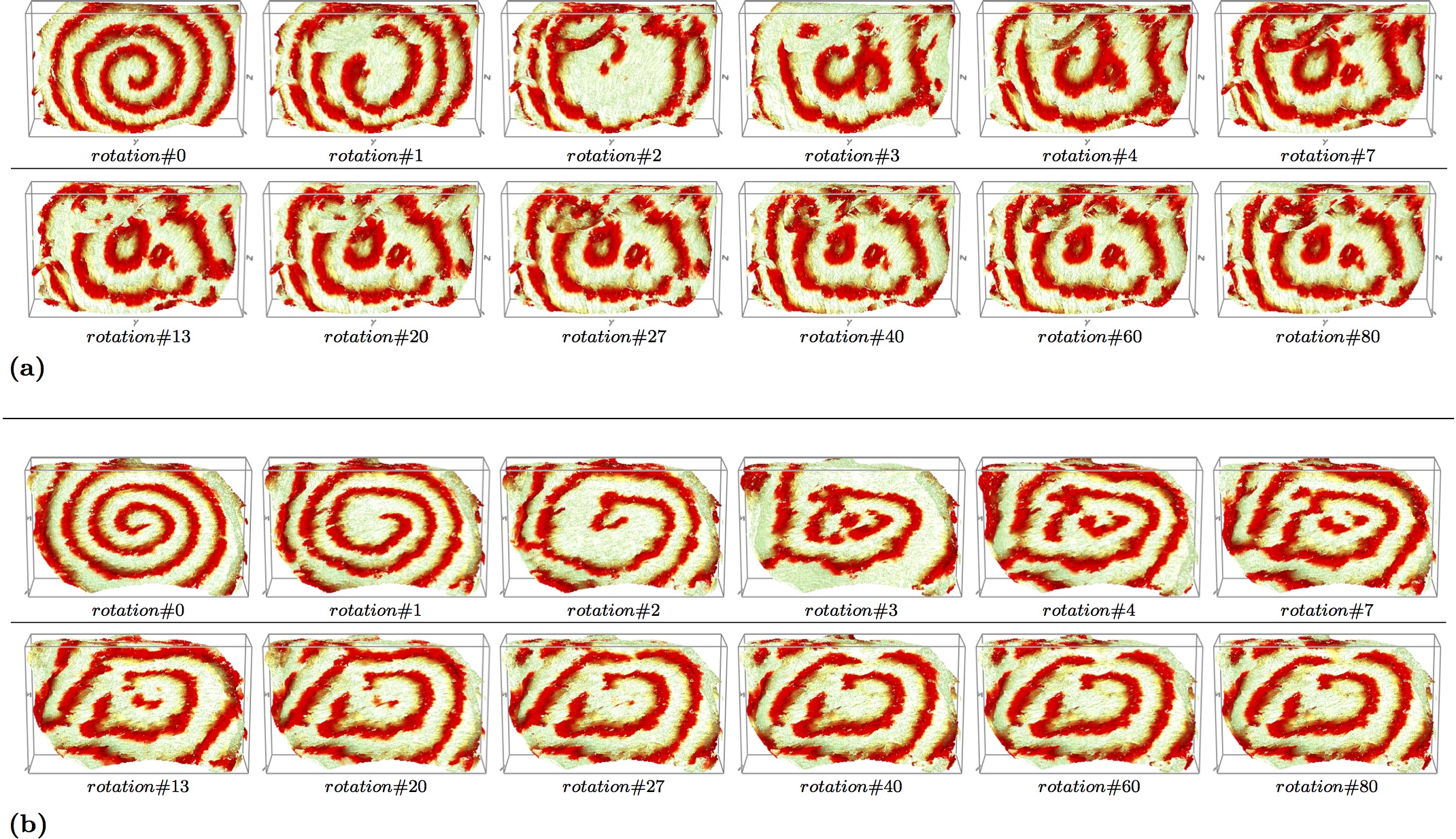}%
\caption{\label{Fig9} {\bf  Rat PV Wall model. Isotropic clockwise transmural re-entry.} The Rat Pulmonary Vein Wall is shown
  in yellow, excitation front shown in red (see the color box in
  Figure~\ref{Fig8}); stroboscopic visualisation
  of re-entry dynamics: time shown under each
  panel as the re-entry rotation number.
The clockwise (in epicardial view) re-entry is initiated transmurally in the middle of the
    PV Wall sample, then fast initial transient is followed by the re-entry
stabilisation. {\bf (a)
  Epicardial view}; (Multimedia view) Fig9a-rev.mpg.
 {\bf (b) Endocardial view}; (Multimedia view) Fig9b-rev.mpg.}
\end{figure*}
\begin{figure*}[tb]
\includegraphics[width=\textwidth]{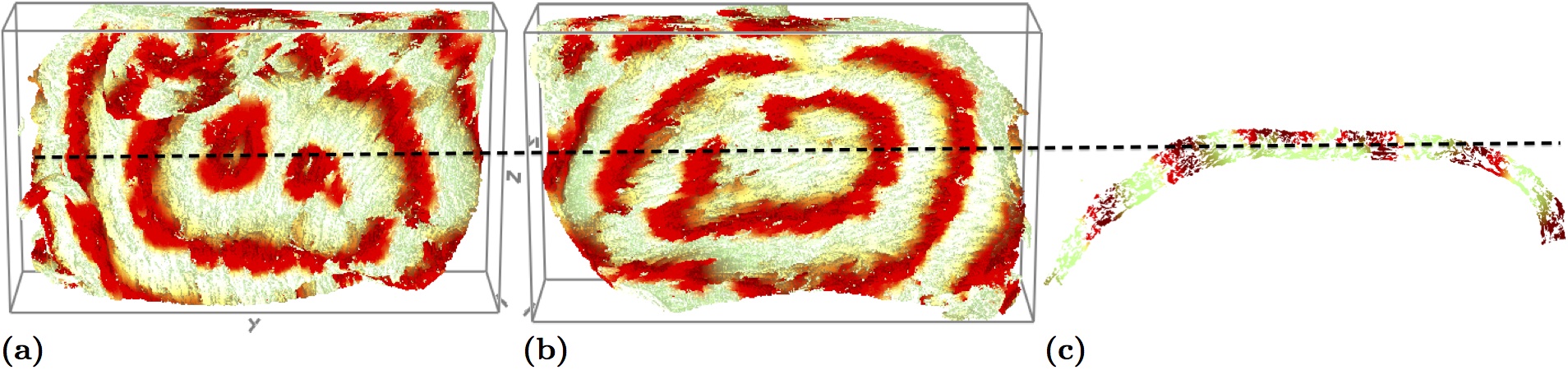}%
\caption{\label{Fig10}{\bf  Rat PV Wall model. Different
      excitation patterns in Epi- and Endo-cardial views}. The Rat Pulmonary Vein Wall is shown
  in yellow, excitation front shown in red (see the color box in
  Figure~\ref{Fig8}; the original isotropic
    clockwise transmural re-entry shown in Figure~\ref{Fig9}. {\bf
      (a)} enlarged \emph{epicardial} view panel ``rotation \# 80'' from
    Figure~\ref{Fig9}(a);  {\bf
      (b)} same ``rotation \# 80'', enlarged \emph{endocardial} view panel from
    Figure~\ref{Fig9}(b); {\bf
      (c)} cross section of the rat PV wall parallel to \code{xy}
    coordinate plane, showing the \emph{transmural structure of the 3D
      excitation pattern} shown in the panels (a-b). The
    dashed line shows alignment of the excitation patterns in all the three panels (a-c). 
}
\end{figure*}
\begin{figure*}[tb]
\includegraphics{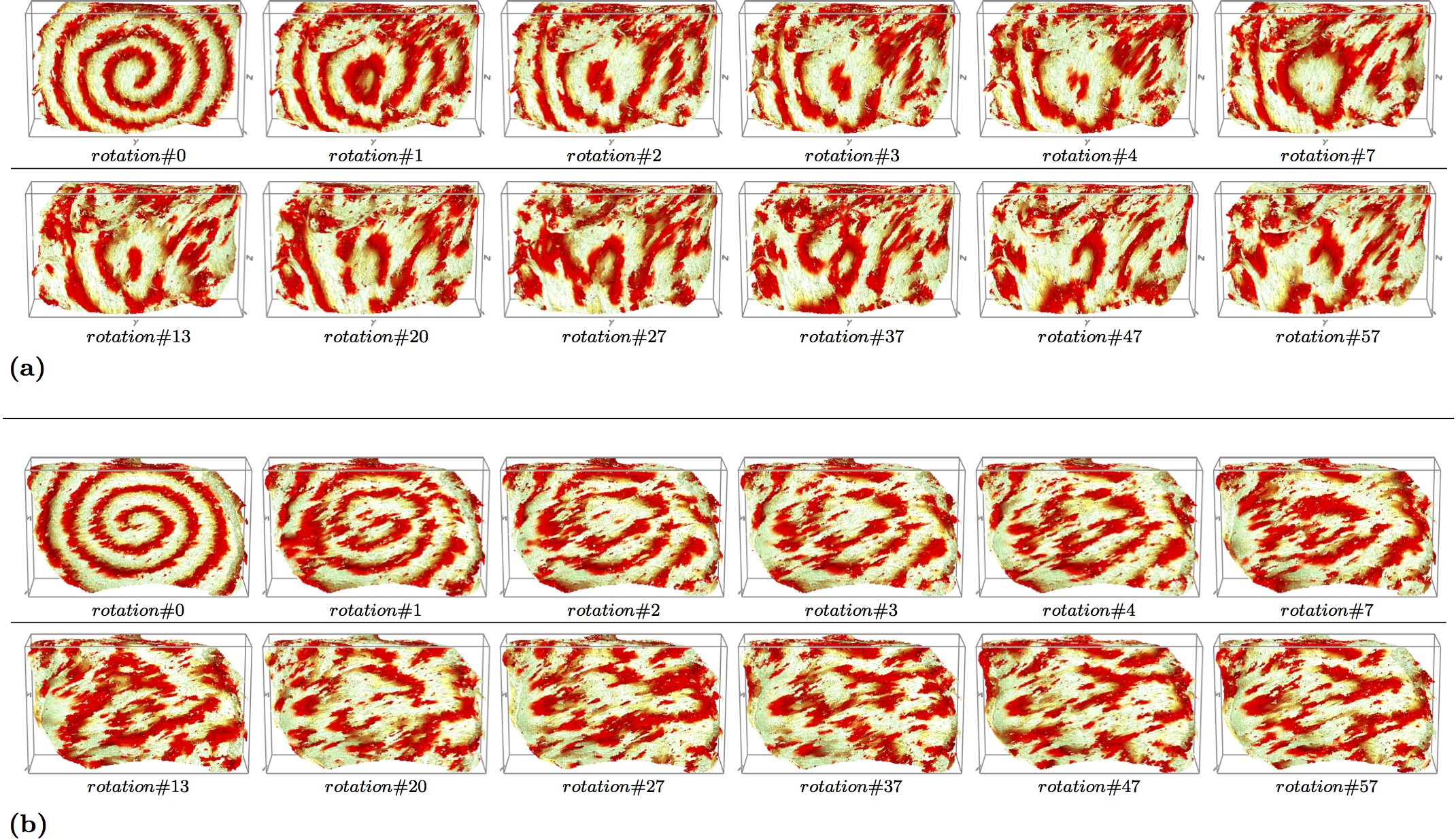}%
\caption{\label{Fig11} {\bf  Rat Pulmonary Vein Wall model. Anisotropic clockwise transmural re-entry.} The Rat Pulmonary Vein Wall is shown
  in yellow, excitation front shown in red (see the color box in
  Figure~\ref{Fig8}); stroboscopic visualisation
  of re-entry dynamics: time shown under each
  panel as the re-entry ``rotation'' number.
The clockwise (in epicardial view) re-entry is initiated transmurally in the middle of the
    PV Wall sample, extremely fast initial transient is followed by
   a chaotic regime without stabilisation. {\bf (a)
  Epicardial view}; (Multimedia view) Fig11a-rev.mpg.
 {\bf (b) Endocardial view}; (Multimedia view) Fig11b-rev.mpg.}
\end{figure*}

The described in Section~\ref{PVWallDataSet} high $3.5\mu m$ subcellular resolution rat PV wall
  micro-CT data set produces quite a ``granular'' PV wall tissue model, see
  Figure~\ref{Fig2}(e,f). So, same as with the sheep atria model
  in Section~\ref{Sheep_results}, in order for a basic test of whether a plane excitation wave propagates through
the granular isotropic PV wall tissue model, a plane excitation wave was
initiated in the PV wall tissue model (Figure~\ref{Fig2}(c,d)), in two perpendicular directions: ``bottom-up'' and
``side-to-side''. 

Figure~\ref{Fig7}(a) shows epicardial and endocardial view of the plane
excitation wave propagating  through
the isotropic PV wall tissue model from the \emph{left to the right} (in the endocardial
view), which proves the connectivity of the tissue model and its
ability to support an excitation wave. An unexpected effect is that the
plane excitation wave's interaction with the PV wall anatomy generates a re-entry pinned to a thickness cusp in the
bottom left (in the endocardial
view,  see also  Figure~\ref{Fig2}(c)) conner of the PV wall, see the 
panels $t=40$,  $t=50$, and $t=150$ in Figure~\ref{Fig7}(a), and (Multimedia view)
Fig7a-rev.mpg. 

Figure~\ref{Fig7}(b) shows epicardial and endocardial views of the plane
excitation wave propagating  through
the isotropic PV wall tissue model 
from the \emph{bottom up} of the PV wall. Interaction of the plane
excitation wave with the epicardial piece of tissue in the form of a ``handle''  in
the top left conner of the PV wall (see also Figure~\ref{Fig2}(d,f)) results
in the break of the plane wave, which can be seen in the epicardial, but not
in the endocardial view panels $t=30$ and $t=40$ in
Figure~\ref{Fig7}(b). The epicardial panel $t=30$
in Figure~\ref{Fig7}(b) shows an ectopic bit going through the
epicardial ``handle'' piece of the tissue, while on the endocardial
side of the PV 
wall nothing indicates the break in the plane wave
propagation. Finally, the interaction of the plane
excitation wave with the epicardial ``handle'' feature results in the
re-entry pinned to this ``handle'' structure, see the 
panels $t=150$ in Figure~\ref{Fig7}(b), and (Multimedia view) Fig7b-rev.mpg.

Figure~\ref{Fig8} shows a stroboscopic epicardial,
Figure~\ref{Fig8}(a), and the corresponding endocardial, Figure~\ref{Fig8}(b),
visualisation of a \emph{counter-clockwise} (in the epicardial view) excitation
vortex initiated in the rat \emph{isotropic} PV wall model by the phase distribution
method~\citep{chaos},  with the initial vortex filament transmurally \emph{along the $x$ axis} through the middle of the wall at $y_0=0.5y_{max},
z_0=0.5z_{max}$. 
From comparison of the epicardial, Figure~\ref{Fig8}(a), and the 
endocardial, Figure~\ref{Fig8}(b), views, it can be seen that
although after a fast anatomy induced transient there is a 
stabilisation of multiple micro-anatomic reentrant drivers, at no point
the epicardial view in Figure~\ref{Fig8}(a) is at least
qualitatively similar to the corresponding instant  
endocardial view in Figure~\ref{Fig8}(b). Rather often what might look a re-entrant driver in
the epicardial view looks a focal source in the
corresponding instant endocardial view, and vice versa.

Figure~\ref{Fig9} shows a stroboscopic epicardial,
Figure~\ref{Fig9}(a), and the corresponding endocardial,
Figure~\ref{Fig9}(b), visualisation of a \emph{clockwise} (in the epicardial view) excitation
vortex initiated in the rat \emph{isotropic} PV wall model by the phase distribution
method~\citep{chaos},  with the initial vortex filament initiated transmurally \emph{along the $x$ axis} through the middle of the wall at $y_0=0.5y_{max},
z_0=0.5z_{max}$. 
Here, after a much faster anatomy induced transient than in the case
of the counter clockwise initial re-entry shown in
Figure~\ref{Fig8}, in Figure~\ref{Fig9}(b) from $\sim rotation\#27$ onwards there is a
stabilisation to what looks like a ``figure eight'' re-entry in the
middle of the instant endocardial views, with the couple of focal
sources in the corresponding 
instant epicardial views in the Figure~\ref{Fig9}(a). 
Again, from comparison of the epicardial,
Figure~\ref{Fig9}(a), and the corresponding
endocardial, Figure~\ref{Fig9}(b), instant views, it can be seen that
although after a fast anatomy induced transient there is a 
stabilisation of multiple micro-anatomic reentrant drivers, at no point
an epicardial instant view in Figure~\ref{Fig9}(a) is
similar to a corresponding
endocardial instant view in Figure~\ref{Fig9}(b). Rather often what might look a re-entrant driver in
the epicardial view looks a focal source in the
corresponding instant endocardial view, and vice versa.

Figure~\ref{Fig10} shows the \emph{transmural structure of the
    instant 3D excitation pattern} seen in the corresponding epicardial
  and endocardial views in Figure~\ref{Fig9} at the instant ``rotation
  \# 80''. Here, Figure~\ref{Fig10}(a) shows the enlarged \emph{epicardial} view panel ``rotation \# 80'' from
    Figure~\ref{Fig9}(a).  Figure~\ref{Fig10}(b) shows the corresponding enlarged \emph{endocardial}
    view panel ``rotation \# 80'' from
    Figure~\ref{Fig9}(b). Figure~\ref{Fig10}(c) shows the
    \emph{transmural structure of the 3D excitation
    pattern} shown in Figure~\ref{Fig10}(a) and Figure~\ref{Fig10}(b),
  in the middle cross section of the rat PV wall parallel to \code{xy}
    coordinate plane. The dashed line shows alignment of all the three panels in 
  Figure~\ref{Fig10}(a-c). From comparison of the \emph{transmural
    structure of the 3D excitation
    pattern} shown in Figure~\ref{Fig10}(c), with the rat PV wall fiber orientation
  structure shown in Figure~\ref{Fig2}(e-f), it can be seen that the
  difference in the PV wall epicardial and endocardial excitation
  patterns correlates with the $\sim 90\degree$ transmural fiber
  rotation. Therefore, there is a clear separation of the ``two focal
  sources''  epicardium excitation pattern from the ``figure of
  eight'' endocardium excitation pattern seen in the cross section view in Figure~\ref{Fig10}(c).

Figure~\ref{Fig11} shows a stroboscopic epicardial,
Figure~\ref{Fig11}(a), and the corresponding endocardial, Figure~\ref{Fig11}(b),
visualisation of a clockwise (in the epicardial view) excitation
vortex initiated in the rat \emph{anisotropic} PV wall model by the phase distribution
method~\citep{chaos},  with the initial vortex filament initiated transmurally \emph{along the $x$ axis} through the middle of the wall at $y_0=0.5y_{max},
z_0=0.5z_{max}$. 
Here, again, no stabilisation of the resulting multiple micro-anatomic
reentrant drivers is achieved, though no termination either,
at least within the simulation time. Same as
  in the corresponding ``isotropic'' case shown in Figure~\ref{Fig9}, at no point  an instant epicardial view in Figure~\ref{Fig11}(a) is similar to the
corresponding instant endocardial view in Figure~\ref{Fig11}(b), with the obvious
predominant excitation propagation along the fiber structures in the
epicardium Figure~\ref{Fig11}(a) at the $\sim90\degree$ to the excitation propagation along the fiber structures in the
endocardium in Figure~\ref{Fig11}(b). 

Note also a \emph{qualitatively different} stabilisation pattern of the
\emph{counter clockwise}, Figure~\ref{Fig8}, and the \emph{clockwise},
Figure~\ref{Fig9}, re-entry in the rat \emph{isotropic} PV wall model simulations.

\section{Discussion}
\label{Discussion}

The role of heart anatomy and anisotropy in the origin
and sustainability of cardiac arrhythmias has been an important open question for a
long time. Anatomy of the heart,
  including its shape, fine anatomical detail, structured anisotropy
  of cardiac myocytes orientation, and the range of transmural
fibre arrangement, shown to be consistent within a species~\citep[p.~173]{Hunter-etal-CompBiolOfHeart},
suggested a possibility of functional role of mammalian
heart anatomy in cardiac re-entry dynamics. The recent discovery of
the new phenomenon of interaction of dissipative
  vortices with small variations of thickness in the
  layer~\citep{Biktasheva-etal-2015-PRL, Ke-etal-Chaos_2015}
  emphasised the possible effects of fine anatomical structures in the heart,
  such as \emph{e.g.} Pectinate Muscles (PM), on the anatomy induced
  drift of cardiac
  re-entry~\citep{Wu-etal-1998-CR,Yamazaki-etal-2012-CVR,Kharche-etal-BioMedRI_2015}. However,
  there is limited experimental
evidence to clarify the exact detail of heart anatomy and anisotropy effect on persistent 
cardiac arrhythmias and fibrillation. 

Combination of High Performance Computing with the
high-resolution DT-MRI, serial histological sections, and micro-CT
data sets directly incorporated into the computationally demanding
complete anatomically realistic computer simulations of the heart 
allows \emph{in-silico}
testing of the heart anatomy and anisotropy effects on the cardiac
re-entry dynamics~\citep{Vigmond-etal-ExperimentalPhysiology-2009,
  Bishop-etal-2010, Kharche-etal-2015-BMRI, Kharche-etal-2015-LNCS,
  bbx-2017-PONE, Biktasheva-etal-HF_2018}.
In this paper, we present anatomy and anisotropy realistic
comparative computer simulation study of cardiac re-entry dynamics in high
resolution isotropic and anisotropic sheep atria models based on the
serial histological sections data sets, Figure~\ref{Fig1}, and in rat pulmonary
vein wall model based on the highest subcellular $3.5\mu m$ resolution micro-CT
data sets, Figure~\ref{Fig2}.
 
In order to eliminate complex effects of realistic cell kinetics, such as \emph{e.g.}
 meander~\citep{Winfree-1991}, alternans~\citep{Karma_Chaos1994}, negative filament tension~\citep{ft},
 etc., and elucidate pure effects of the heart anatomy and anisotropy
 on cardiac re-entry dynamics, simplified
 FitzHugh-Nagumo~\citep{Winfree-1991} excitation model with rigidly
 rotating core and positive filament tension was chosen for this
 study.  A specific interplay of
 realistic cardiac excitation kinetics with heart anatomy
 and anisotropy, such as e.g. frequency lock, should be addressed in future studies. 

The dominant periods of all the simulated re-entries in the ``down-sampled'', Figure~\ref{Fig1}(c) and Figure~\ref{Fig3}(c), and in the 
``complete'', Figure~\ref{Fig1}(a,b) and Figure~\ref{Fig5}, sheep
atria models were close to the period 11.36 of the free spiral wave in the
FitzHugh-Nagumo model (see Section~\ref{RD}), which was expected as we considered homogeneous
models without any strong perturbations, and also served as an additional
confirmation that we were dealing with functional rather than anatomical re-entries. 

Our by all means initial comparative isotropic and anisotropic simulation of cardiac
re-entry in the otherwise homogeneous high resolution models of sheep atria confirmed
cardiac re-entry interaction with the fine anatomical
structures~\citep{Wu-etal-1998-CR,Yamazaki-etal-2012-CVR,Kharche-etal-2015-BMRI,
Biktasheva-etal-2015-PRL}. It can be seen in Figure~\ref{Fig3}(b,d)
and Figure~\ref{Fig4}, that the abundance of fine endocardial
structures in the sheep atria provides a lot of potential pinning sites for
sheep atrial re-entry. So, depending on the re-entry initial location,
it could pin to Bachman’s
bundle and pectinate muscles in the left and right atrial appendages, see
Figure~\ref{Fig3}(b,d) and Figure~\ref{Fig4}. This pinning to fluctualtions
of thickness in the sheep atria can be seen in the less accurate ``down-sampled'' model, Figure~\ref{Fig1}(c) and Figure~\ref{Fig3}(b,c), as well as in the high
resolution ``complete'' sheep atria model, Figure~\ref{Fig1}(a, b),
Figure~\ref{Fig3}(d), Figure~\ref{Fig4} and Figure~\ref{Fig5}.

From Figure~\ref{Fig4} and Figure~\ref{Fig5}, it can be seen that, in
general, anisotropy significantly changes the trajectory and pinning
location of the anatomy induced drift. A comparison of the
trajectories ``with'' (trajectories ``(a)'', ``(c)'', ``(e)'', and
``(g)'' in Figure~\ref{Fig4}) and ``without'' anisotropy (trajectories ``(b)'', ``(d)'', ``(f)'', and
``(h)'' in Figure~\ref{Fig4}), suggests that the sheep atria anisotropy
tends to drift a re-entry away from the main blood vessesls into the
right and left atrial appendages, although, of course, such a
generalisation would require a lot more simulations in order to be
statistically salient.

Theory~\citep{Wellner-etal-PNAS2002, Dierckx-etal-2015-PRL} predicts strong effects of anisotropy on dynamics of singular filaments, which must act in
        addition to purely geometric forces.
An iteresting confirmation of the theoretical prediction is the case of isotropic vs
    anisotropic conductance in the sheep atria, which resulted in the 
    very different lengths and orientation of the final pinned
    filaments shown in Figure~\ref{Fig6}(b) and in
    Figure~\ref{Fig6}(c). Different to the short \emph{isotropic transmural} filament shown in
    Figure~\ref{Fig6}(b), the realistic anysotropy of the very thin
    sheep atria still was able to result in a long \emph{intramural} filament stretched almost parallel to the epicardial surface,
    Figure~\ref{Fig6}(c). The transient increase in total number and
    length of re-entry filaments was shown to be a mechanism of
    re-entry self-termination in the DT-MRI based model of human foetal
    heart~\citep{Biktasheva-etal-HF_2018}. Although we also observed
    here the increase of the length of sheep atrial re-entry filaments
    due to the sheep atria anisotropy, we have not seen in the present
    sheep atria
    study a transient increase in
    number and length of the filaments followed by the
    self-termination of the re-entry, as e.g. in human foetal heart simulations~\citep{Biktasheva-etal-HF_2018}. 

A comparison of the present study of anatomy induced drift in sheep atria with our previous studies of re-entry dynamics in
human atria~\citep{Kharche-etal-CinC2012, Kharche-etal-HA_IEEE2013,
  Kharche-etal-FIMH_2015, Kharche-etal-2015-BMRI}, also suggests that
different heart anatomy in 
different species can result in different patterns
of atrial re-entry dynamics, albeit the general tendency to pin to a
nearest stable pinning location persists~\citep{Biktashev-etal-2011-PONE}.

The comparative simulations of micro-anatomic re-entry in
the highest $3.5\mu m$ subcellular resolution rat PV wall model, where the transmural fiber direction
change of $\sim90\degree$ is a salient feature of the PV wall
anatomy, see Figure\ref{Fig2}(e,f), gave a number of
important confirmations and new unexpected results.

Note that at the subcellular $3.5\mu m$  spatial resolution used, despite the obvious limitation due to the lack of biophysical
realism, there are
meaningful observations to be made. Namely, at this resolution, the
anisotropic structure of the tissue is represented directly, see
Figure~\ref{Fig2}(c,d), unlike
the averaged representation in more traditional macroscopic
models. 
Consequently, following the convenient alignment of the visible due to the high subcellular resolution
orientation of the fibres in the rat PV wall micro-CT
images, that is the \emph{endocardium} fibers were aligned with the
horisontal \code{x} axis (Figure~\ref{Fig2}(c)), while the \emph{epicardium}
fibers were aligned with the vertical \code{y} axis (Figure~\ref{Fig2}(d)), 
the effect of the fibre directions is observed
in simulations with scalar (isotropic) diffusivity. 

The plane wave simulation shown in Figure~\ref{Fig7}, confirmed a
connectivity and capability of the subcellular resolution rat PV wall
tissue model to propagate excitation waves. It also showed that the
specific combination of the rat PV wall anatomy, including the small
fluctuations of thickness due to the fine endocardium and epicardium
structural features, together with the transmural fiber direction
change of $\sim90\degree$, Figure~\ref{Fig2}(c-f), may generate a re-entry out of a plane
excitation wave without any electrophysiological
inhomogeneities. Various ``geometric'' mechanisms of re-entry initiation have been
  discussed by theoreticians~\citep{Agladze-etal-Science_1994,
    ZemlinPertsov-Europace_2007}; we stress though that in our case, whichever
  mechanism played the decisive role, the geometry was not in any way specially
  prepared but obtained from a real tissue preparation, and no specific stimulation
  protocol was applied apart from a single plane wave initiation.  

From simulations shown in Figure~\ref{Fig8} and Figure~\ref{Fig9}, it
can be seen that re-entry of different chirality
produce qualitatively different stabilisation patterns in the rat PV
wall tissue model.

Importantly, the simulations confirmed that, because of the transmural fiber direction
change of $\sim90\degree$ in the PV wall, a stable atrial micro-anatomic re-entry
might produce qualitatively different endocardial and epicardial
manifestation~\citep{Csepe-etal-2017}, so that a clearly seen on the
endocardium micro-anatomic ``figure eight'' reentry,
Figure~\ref{Fig10}(b), might appear as a couple of focal
sources on the epicardium, Figure~\ref{Fig10}(b, c), and vice
  versa: compare \emph{e.g.} the corresponding time instant panels
  ``rotation \# 6'' in Figure~\ref{Fig8}(a) and in
  Figure~\ref{Fig8}(b). Appreciation of the fact that even a rather thin
  the PV wall may still produce an essentially 3D excitation
  patten, as e.g. shown  in
  Figure~\ref{Fig10}, might be important in atrial ablation when \emph{e.g.} determining location of a
  re-entrant circuit.

 Introduction of a diffusivity tensor based on the PV wall fibre directions further
enhances the effect of anisotropy in the PV wall model. See \emph{e.g.}
Figure~\ref{Fig11}, which shows the expected enhanced \emph{epicardium}
propagation of excitation along the vertical \code{y} axis, and the enhanced \emph{endocardium}
propagation of excitation along the horisontal \code{x} axis, but in this context such
enhancement might be entirely artificial.

Although a general role of fiber anisotropy in the genesis and sustenance of
arrhythmias could be and has been addressed by numerics in idealised and
simplified geometries with different spatial distributions of
anisotropy, see e.g. the pioneering study~\citep{Fenton-Karma-1998},  we believe that the main novelty and
significance of our present study
is that the high resolution serial histological sections and
the subcellular resolution
micro-CT data sets, containing both the detail heart anatomy and the
myofiber structure, can be directly incorporated into the
computationally demanding HPC complete anatomy and anisotropy
realistic computer simulations, and thus serve a model for comparative intra- and
inter-species in-silico study of cardiac re-entry dynamics in mammalian hearts.

\section{Nomenclature}

\subsection{Resource Identification Initiative}
\bbx , RRID:SCR\_015780

\section*{Conflict of Interest Statement}

The authors declare that the research was conducted in the absence of
any commercial or financial relationships that could be construed as a
potential conflict of interest.

\section*{Animal ethics statement}

All surgical procedures were approved by the Animal Ethics Committee of the University of Auckland and conform to the Guide for the Care and Use of Laboratory Animals (NIH publication no. 85-23)

\section*{Author Contributions}

GSR, BT, IJLeG, BHS, JZ, and IVB contributed conception and design of the study; GSR, BT,
IJLeG, BHS, and JZ contributed the micro-CT and serial histological
sections data sets; BT wrote Section~\ref{Sheep-DataSet}; GSR wrote Section~\ref{PVWall-microCT}; GSR, BT, and VNB
converted the data sets into \bbx \; 
geometry format; IVB, VNB, and DPF ran the
simulations; IVB and VNB did visualisation; IVB performed the 
analysis, and wrote the first draft of the manuscript.  All authors contributed to manuscript
revision, read and approved the submitted version.

\section*{Funding}
We acknowledge support of the Health Research Council of New Zealand
for the data sets. 
Development of \bbx \, software was supported by EPSRC (UK) grant EP/I029664.
We also acknowledge partial support by EPSRC (UK) grants
EP/E016391/1, EP/N014391/1, EP/E018548/1, EP/P008690/1; by the University of Sheffield (UK) Network:
POEMS - Predictive mOdelling for hEalthcare technologies through MathS under EPSRC (UK) grant EP/L001101/1; by the National Science Foundation (USA)
under Grants No. NSF PHY99-07949 and NSF PHY-1748958.

\section*{Acknowledgements}
We thank all the developers of the BeatBox HPC Simulation Environment
for Biophysically and Anatomically Realistic Cardiac
Electrophysiology. 
We thank EPSRC Centre for Predictive Modelling in Healthcare,
University of Exeter, Exeter, UK  for running the simulations on the CPMH compute servers.

\section*{Supplementary Material}
\label{Suppl}
The Supplementary Material for this article can be found online at ...

\bibliographystyle{frontiersinSCNS_ENG_HUMS} 


\end{document}